\documentclass[10pt,twocolumn]{IEEEtran}

\usepackage{cite}
\usepackage{amsmath,amssymb,amsfonts}
\usepackage{graphicx}
\usepackage{booktabs}
\usepackage{multirow}
\usepackage{array}
\usepackage{subcaption}
\usepackage{float}
\usepackage{url}
\usepackage{hyperref}
\usepackage{xcolor}

\hypersetup{
    colorlinks=true,
    linkcolor=blue,
    citecolor=blue,
    urlcolor=blue
}

\usepackage{cite}

\usepackage{tikz}
\usetikzlibrary{positioning,arrows.meta}
\usetikzlibrary{shapes.geometric}

\begin{document}

\title{Spectrum Coexistence, Network Dimensioning, and Cell-Free Architectures in 5G and 5G-Advanced Wireless Networks}

\author{
Siminfar~Samakoush~Galougah
\thanks{S. S. Galougah is with the Department of Electrical and Computer Engineering, University of Maryland, College Park, USA (e-mail: simin95@umd.edu).}
}

\maketitle

\begin{abstract}
Fifth-generation (5G) wireless networks introduce new architectural paradigms, spectrum usage models, and optimization challenges to support enhanced mobile broadband, massive machine-type communications, and ultra-reliable low-latency communications.
This survey provides a comprehensive overview of key technologies and design challenges in 5G systems, with a focus on spectrum coexistence and interference management, network dimensioning and planning, cell-free massive MIMO architectures, fronthaul-aware user management, and power allocation strategies.
Representative analytical, simulation-based, and optimization-driven approaches are reviewed, fundamental trade-offs are highlighted, and open research challenges relevant to 5G-Advanced and beyond are identified.
\end{abstract}


\section{Introduction}
\label{sec:introduction}

The fifth generation (5G) of wireless networks has been designed to meet unprecedented performance requirements in terms of peak data rate, latency, reliability, and connectivity density, enabling services such as enhanced mobile broadband (eMBB), massive machine-type communications (mMTC), and ultra-reliable low-latency communications (URLLC)~\cite{b01,b16}.
To support these heterogeneous services, 5G departs significantly from previous generations by introducing flexible spectrum usage, heterogeneous deployment scenarios, and highly distributed radio access network (RAN) architectures.

In contrast to traditional cell-centric designs, 5G networks operate across a wide range of frequency bands, including low-band, mid-band, and millimeter-wave (mmWave) spectrum, and across licensed, unlicensed, and shared regulatory regimes.
At the same time, architectural paradigms such as centralized and cloud-native RAN, network function virtualization, and cell-free massive MIMO fundamentally change how radio resources are coordinated and optimized~\cite{b16}.
While these innovations greatly enhance flexibility and scalability, they also give rise to new technical challenges that span spectrum coexistence and interference management, network dimensioning and planning, distributed signal processing, and fronthaul-aware resource optimization.

\subsection{Motivation}

A substantial body of literature has investigated individual aspects of 5G systems, including spectrum coexistence mechanisms~\cite{b1,b7}, stochastic-geometry-based network dimensioning~\cite{b10,b11}, and cell-free massive MIMO architectures~\cite{001b1,001b7}.
However, much of the existing work treats these topics in isolation, focusing on specific layers, deployment scenarios, or optimization objectives.
In practical 5G deployments, these aspects are tightly coupled: spectrum coexistence constraints directly affect interference statistics and resource availability; network dimensioning depends on spatial user distributions and spectrum access policies; and distributed architectures such as cell-free massive MIMO introduce fronthaul constraints that fundamentally shape user association and power allocation strategies.

Moreover, recent trends toward spectrum sharing, dense heterogeneous deployments, and user-centric architectures further blur the boundaries between classical planning, physical-layer design, and system-level optimization.
As a result, there is a growing need for a unified survey that connects spectrum-level considerations with architectural choices and optimization techniques, rather than addressing each component in isolation.
Such a holistic perspective is particularly relevant as 5G evolves toward 5G-Advanced, where tighter integration of radio access, transport, and compute resources is expected to play a central role.

This survey aims to fill this gap by providing a coherent overview of key 5G technologies and challenges, with emphasis on how spectrum coexistence, stochastic network dimensioning, cell-free massive MIMO architectures, and fronthaul-aware power and resource optimization interact in realistic deployment scenarios.

\subsection{Contributions}

The main contributions of this survey are summarized as follows:
\begin{itemize}
    \item A structured overview of major 5G architectural paradigms and spectrum usage models, highlighting their implications for interference management and system design.
    \item A unified taxonomy of spectrum coexistence mechanisms, stochastic-geometry-based network dimensioning methods, and cell-free massive MIMO architectures.
    \item A comparative discussion of resource optimization strategies, including user management and power control, under fronthaul and interference constraints.
    \item Identification of open research challenges and emerging directions relevant to 5G-Advanced and future wireless systems.
\end{itemize}
\begin{figure}[t]
\centering
\begin{tikzpicture}[
    block/.style={rectangle, draw, rounded corners,
                   align=center, minimum width=3.3cm,
                   minimum height=1.1cm},
    arrow/.style={->, thick}
]

\node[block] (spec) {Spectrum \& Regulation\\
{\footnotesize Licensed / Shared / Unlicensed}};
\node[block, below=0.8cm of spec] (coex) {Coexistence \& Interference\\
{\footnotesize ACIR, TDD alignment}};
\node[block, below=0.8cm of coex] (geo) {Spatial Statistics\\
{\footnotesize User PPPs, ISD}};
\node[block, below=0.8cm of geo] (dim) {Network Dimensioning\\
{\footnotesize Overload probability}};
\node[block, below=0.8cm of dim] (arch) {Distributed Architecture\\
{\footnotesize Cell-free / User-centric}};
\node[block, below=0.8cm of arch] (opt) {Fronthaul \& Optimization\\
{\footnotesize User mgmt, Power control}};

\draw[arrow] (spec) -- (coex);
\draw[arrow] (coex) -- (geo);
\draw[arrow] (geo) -- (dim);
\draw[arrow] (dim) -- (arch);
\draw[arrow] (arch) -- (opt);

\end{tikzpicture}
\caption{Cross-layer structure of the survey, highlighting the coupling between spectrum coexistence, spatial network modeling, architectural design, and fronthaul-aware resource optimization in 5G and 5G-Advanced networks.}
\label{fig:overview}
\end{figure}
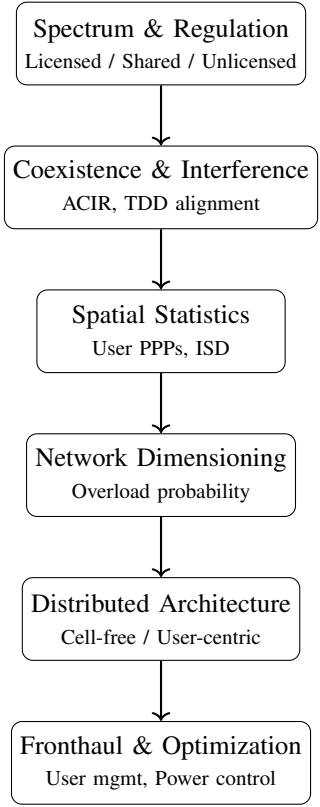

\subsection{Organization}

The remainder of this paper is organized as follows.
Section~\ref{sec:architecture} reviews 5G network architecture and spectrum allocation frameworks.
Section~\ref{sec:coexistence} discusses spectrum coexistence and interference management in heterogeneous 5G deployments.
Section~\ref{sec:dimensioning} focuses on network dimensioning and planning using stochastic geometry.
Section~\ref{sec:cellfree} reviews cell-free massive MIMO systems and fronthaul-aware user management strategies.
Section~\ref{sec:power} addresses power allocation and resource optimization under practical constraints.
Section~\ref{sec:future} outlines open challenges and future research directions, and Section~\ref{sec:conclusion} concludes the paper.

\section{5G Architecture and Spectrum Overview}
\label{sec:architecture}

\subsection{5G Network Architecture}

Unlike previous generations, fifth-generation (5G) networks adopt a highly flexible and software-driven architecture to support heterogeneous service requirements, including enhanced mobile broadband (eMBB), massive machine-type communications (mMTC), and ultra-reliable low-latency communications (URLLC)~\cite{b01,b16}.
To meet these diverse requirements, 5G evolves from rigid, cell-centric designs toward more disaggregated, virtualized, and cloud-native architectures.

\paragraph{Centralized architectures.}
Centralized radio access network (C-RAN) architectures pool baseband processing at centralized units (CUs), while remote radio heads (RRHs) are deployed at cell sites.
This architectural model enables efficient coordination, interference management, and resource pooling across cells~\cite{b16}.
However, C-RAN imposes stringent fronthaul requirements in terms of capacity and latency, which can become a limiting factor in dense deployments or wideband operation.

\paragraph{Distributed architectures.}
To alleviate fronthaul bottlenecks, 5G introduces flexible functional splits between centralized and distributed units (DUs), allowing parts of the physical layer to be executed closer to the radio units~\cite{b37}.
This distributed RAN (D-RAN) approach provides a trade-off between coordination gains and transport constraints, enabling scalable deployments across urban, suburban, and indoor scenarios.

\paragraph{Cloud-native and virtualized architectures.}
A defining characteristic of 5G is the adoption of network function virtualization (NFV) and software-defined networking (SDN), whereby radio access and core network functions are implemented as virtual network functions running on general-purpose hardware~\cite{b01,b16}.
More recent cloud-native implementations further decompose network functions into microservices, enabling elastic scaling and rapid service provisioning.
These architectural trends enable advanced capabilities such as network slicing, in which multiple logical networks with distinct performance guarantees coexist on shared physical infrastructure~\cite{b27}.

Overall, the architectural evolution of 5G reflects a shift toward programmability and flexibility, at the cost of increased system complexity and tighter coupling between radio design, transport constraints, and resource optimization recurring themes throughout this survey.

\subsection{Spectrum Allocation and Deployment Scenarios}
Spectrum availability and utilization fundamentally shape 5G system performance.
To accommodate diverse coverage and capacity requirements, 5G operates across a wide range of frequency bands and regulatory regimes, resulting in heterogeneous deployment scenarios~\cite{b01,b16}.

\paragraph{Low-band spectrum ($<1$\,GHz).}
Low-frequency bands offer wide-area coverage and favorable propagation characteristics, making them suitable for mMTC services and rural deployments.
However, limited available bandwidth restricts achievable data rates, motivating their primary use for coverage-oriented services~\cite{b16}.

\paragraph{Mid-band spectrum (1--6\,GHz).}
Mid-band frequencies, including sub-6\,GHz allocations such as the 3.3--3.8\,GHz range, provide a balance between coverage and capacity.
These bands are widely regarded as the backbone of 5G deployments, supporting both macro-cell and dense urban scenarios~\cite{b01,b27}.
Due to increasing congestion, mid-band operation frequently involves spectrum sharing and coexistence with incumbent or adjacent systems.

\paragraph{Millimeter-wave (mmWave) spectrum ($>24$\,GHz).}
Millimeter-wave (mmWave) bands offer large contiguous bandwidths that enable multi-gigabit data rates for eMBB applications~\cite{b5}.
However, their propagation characteristics lead to limited coverage, strong sensitivity to blockage, and the need for highly directional beamforming.
As a result, mmWave deployments typically rely on dense small-cell layouts and advanced antenna technologies.

\paragraph{Licensed, unlicensed, and shared spectrum.}
5G supports operation in licensed, unlicensed, and shared spectrum regimes.
Licensed spectrum provides predictable interference conditions but is scarce and costly.
Unlicensed and shared spectrum improve spectral efficiency but introduce coexistence and interference-management challenges~\cite{b7,b20}.
Regulatory frameworks such as dynamic spectrum access and spectrum commons require mechanisms including power control, time-domain coordination, and emission constraints to ensure fair coexistence among heterogeneous systems.

\paragraph{Deployment heterogeneity.}
In practical deployments, 5G networks combine multiple frequency bands, spectrum regimes, and architectural configurations.
Outdoor macro cells, indoor small cells, and mixed outdoor--indoor scenarios often coexist, leading to complex interference patterns and planning challenges~\cite{b1}.
This deployment heterogeneity directly motivates the coexistence analysis, stochastic dimensioning, and distributed optimization techniques reviewed in the subsequent sections of this survey.

\section{Spectrum Coexistence and Interference Management in 5G}
\label{sec:coexistence}
Spectrum scarcity, heterogeneous deployments, and the introduction of new mid- and high-band spectrum have made coexistence a central design axis in 5G/NR, particularly when multiple operators or systems transmit in \emph{adjacent channels}~\cite{b01,b1,b2,b7,b16,b27}. In adjacent-band coexistence, interference is driven by (i) \emph{transmitter leakage} (out-of-band emission) and (ii) \emph{receiver selectivity} (imperfect rejection of adjacent-band energy)~\cite{b22,b24,b33}. These effects jointly determine the effective adjacent-channel interference coupling between an \emph{aggressor} system and a \emph{victim} system, and directly influence regulatory emission masks, spectrum-sharing frameworks, and deployment guidelines defined for 5G and IMT-2020 systems~\cite{b25,b26,b37}.

\subsection{Leakage, Selectivity, and ACIR}
\paragraph{Transmitter leakage (ACLR).}
Non-ideal transmit filtering and RF front-end non-linearities cause part of the transmitted power to leak outside the allocated channel, producing out-of-band emissions (OOBE) that interfere with neighboring systems~\cite{b24,b33}. A standard metric to characterize this effect is the adjacent channel leakage ratio (ACLR),
\begin{equation}
\mathrm{ACLR} \triangleq \frac{P_{\mathrm{Tx}}}{P_{\mathrm{OOBE}}},
\end{equation}
where $P_{\mathrm{Tx}}$ is the in-channel transmit power and $P_{\mathrm{OOBE}}$ is the power leaked into the adjacent channel bandwidth. ACLR limits and emission masks are explicitly specified by 3GPP for LTE and NR base stations and user equipment across frequency ranges and deployment scenarios~\cite{b37,b25}.

\paragraph{Receiver selectivity (ACS).}
Finite receiver filter steepness leads to imperfect suppression of adjacent-band transmissions, quantified by the adjacent channel selectivity (ACS)~\cite{b22,b24},
\begin{equation}
\mathrm{ACS} \triangleq \frac{P_{\mathrm{Rx}}}{P_{\mathrm{OOB}}},
\end{equation}
where $P_{\mathrm{OOB}}$ denotes the received power passing through the receiver filter from outside the desired channel. While receiver requirements are generally less stringent than transmitter emission limits, ACS plays a key role in interoperability and coexistence performance, particularly in dense deployments~\cite{b22,b37}.

\paragraph{Combined coupling (ACIR).}
The net adjacent-channel interference coupling is captured by the adjacent channel interference ratio (ACIR), which combines ACLR and ACS (in linear scale)~\cite{b22}:
\begin{equation}
\mathrm{ACIR} \triangleq \left(\frac{1}{\mathrm{ACLR}}+\frac{1}{\mathrm{ACS}}\right)^{-1}.
\end{equation}
ACIR provides a compact abstraction that links RF emission and selectivity specifications to system-level interference and SINR outcomes, and is widely used in coexistence analysis and spectrum-sharing studies~\cite{b22,b24,b33}.

\subsection{TDD Frame Configuration and Cross-Link Interference}
In TDD multi-cell deployments, coexistence depends strongly on whether neighboring systems align their DL/UL subframes. When frames are \emph{aligned} (e.g., UL--UL), interference is typically dominated by UE-to-BS coupling; when frames are \emph{conflicting} (e.g., victim UL while aggressor DL), strong BS-to-BS or BS-to-UE cross-link interference can occur~\cite{b34,b38}. Standard bodies and operators focus on BS desensitization because BS transmit powers significantly exceed UE transmit powers, and BS outages can simultaneously impact many users. As a result, regional frame-structure coordination and traffic-load-aware alignment strategies have been proposed, albeit at the cost of reduced flexibility in UL/DL adaptation~\cite{b34,b38}.

\begin{figure}[t]
\centering
\begin{tikzpicture}[
    bs/.style={rectangle, draw, minimum width=0.95cm, minimum height=0.95cm, align=center},
    ue/.style={circle, draw, minimum size=0.65cm, align=center},
    arrowUL/.style={->, thick},
    arrowDL/.style={->, thick, dashed},
    title/.style={font=\bfseries},
    note/.style={font=\scriptsize, align=center}
]

\def\xL{-2.0}
\def\xR{ 2.0}
\def\yT{2.05}
\def\yB{1.0}
\def\dx{1.0}

\node[title] at (\xL,\yT) {Aligned UL--UL};
\node[note]  at (\xL,1.65) {UE$\rightarrow$BS dominates};
\node[bs] (bs1) at (\xL-\dx,\yB) {BS};
\node[ue] (ue1) at (\xL+\dx,0.0) {UE};
\draw[arrowDL] (ue1) -- (bs1);

\node[title] at (\xR,\yT) {Conflicting UL--DL};
\node[note]  at (\xR,1.65) {BS$\rightarrow$BS dominates};
\node[bs] (bs2) at (\xR-\dx,\yB) {BS};
\node[bs] (bs3) at (\xR+\dx,\yB) {BS};
\draw[arrowDL] (bs3) -- (bs2);

\node[note] (call) at (\xR,0.15) {strong cross-link\\interference};

\end{tikzpicture}
\caption{Conceptual illustration of adjacent-band coexistence under TDD operation. Aligned UL--UL frames are dominated by UE-driven interference, while conflicting UL--DL frames lead to strong BS-driven cross-link interference.}
\label{fig:tdd}
\end{figure}
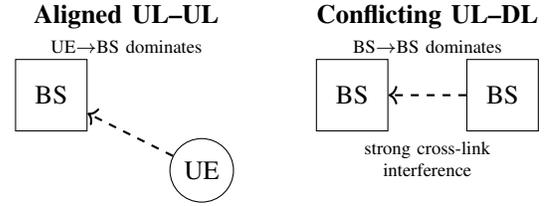

\subsection{Case Studies in CBRS Adjacent Bands (3.5--3.7\,GHz)}
To illustrate practical coexistence behavior, two baseline scenarios are evaluated for adjacent-band operation in the CBRS range, a representative shared mid-band spectrum with regulatory coexistence constraints~\cite{b17}. The victim system is configured in uplink to study BS desensitization, while the aggressor alternates between downlink (conflicting frames) and uplink (aligned frames). Such scenarios are consistent with coexistence testing methodologies and deployment studies for shared-spectrum and licensed mid-band NR systems~\cite{b20,b25,b37}.

\subsubsection{Outdoor--Outdoor (Macro) Networks}
A multi-cell scenario with $7$ victim outdoor macro BSs and $19$ aggressor macro BSs is considered, each BS split into three sectors with multiple UEs per sector. Two inter-site distances (ISD) are examined: $100$\,m and $300$\,m, reflecting dense urban and more relaxed macro deployments~\cite{b17}.

A consistent observation is that \textbf{conflicting frames increase interference and reduce SINR} relative to aligned frames, in agreement with prior studies on TDD cross-link interference~\cite{b17}. For example:
\begin{itemize}
\item At $\mathrm{ISD}=100$\,m, the reported average interference increases from approximately $-113.1$\,dBm (aligned UL--UL) to $-104.7$\,dBm (conflicting UL--DL), with average SINR decreasing from about $6.2$\,dB to $4.6$\,dB.
\item At $\mathrm{ISD}=300$\,m, the reported average interference increases from approximately $-123.7$\,dBm (aligned UL--UL) to $-107.8$\,dBm (conflicting UL--DL), with average SINR decreasing from about $6.11$\,dB to $5.64$\,dB.
\end{itemize}
These trends reflect the combined effects of higher BS transmit power relative to UE power and increased path loss at larger ISD, which jointly shape adjacent-band interference coupling~\cite{b17}.

\subsubsection{Outdoor Aggressor -- Indoor Victim}
A mixed scenario with one outdoor macro aggressor and one indoor micro victim is studied with $120$\,m separation. Indoor penetration loss is modeled via wall attenuation, effectively increasing isolation between aggressor and victim depending on the interfering link type~\cite{b17}.

Across TDD frame combinations (DL--DL, DL--UL, UL--UL, UL--DL), the strongest interference occurs in the \textbf{UL--DL conflicting case}, consistent with BS-driven cross-link interference being the dominant impairment even in the presence of indoor isolation~\cite{b34,b38}. In contrast, aligned UL--UL yields substantially lower interference due to UE-limited transmit power and additional attenuation from building penetration~\cite{b20}.

\subsection{Design Implications for Adjacent-Band Coexistence}
From a survey perspective, this study reinforces several widely accepted coexistence insights:
\begin{enumerate}
\item \textbf{Frame alignment is a powerful coexistence lever} in adjacent-band TDD deployments, particularly to mitigate BS-to-BS interference~\cite{b34,b38,b17}.
\item \textbf{Deployment geometry matters:} reduced ISD can significantly amplify adjacent-band interference under conflicting frames~\cite{b16,b34,b17}.
\item \textbf{Outdoor--indoor isolation can help,} but does not fully prevent severe interference in BS-driven conflicting configurations~\cite{b20,b26,b17}.
\item \textbf{ACIR-based abstractions bridge RF and system design,} enabling translation from emission/selectivity requirements to SINR and throughput impacts~\cite{b22,b24,b33,b37}.
\end{enumerate}

\begin{table}[t]
\centering
\caption{Headline adjacent-band coexistence trends under TDD frame alignment vs conflict.}
\label{tab:coexistence_trends}
\begin{tabular}{p{2.1cm}p{2.2cm}p{3.2cm}}
\toprule
\textbf{Scenario} & \textbf{Frames} & \textbf{Observed trend} \\
\midrule
Outdoor--Outdoor & UL--UL (aligned) & Lower interference, higher SINR (UE-driven interference) \\
Outdoor--Outdoor & UL--DL (conflict) & Higher interference, lower SINR (BS-driven interference) \\
Outdoor--Indoor & UL--DL (conflict) & Worst case; BS-to-(indoor) victim dominates despite wall loss \\
Outdoor--Indoor & UL--UL (aligned) & Best case; UE-limited interference and indoor isolation \\
\bottomrule
\end{tabular}
\end{table}

\section{Network Dimensioning and Planning in 5G}
\label{sec:dimensioning}

A key planning problem in OFDMA-based cellular systems is \emph{dimensioning} the number of available subcarriers (or resource units) so that the probability of \emph{resource shortage} remains below a target threshold~\cite{b0025,b0020}. This problem has long been studied in earlier generations of cellular networks using traffic models, Erlang-based formulations, and optimization-based planning tools~\cite{b001,b002,b008,b009,b0019}. It becomes particularly important in IoT settings, where a massive number of low-to-moderate rate devices with heterogeneous quality-of-service requirements compete for limited time--frequency resources~\cite{b0018,b0021}. 

A principled modern approach is to explicitly model user spatial randomness and map it to random resource demand, then analyze or bound the overload probability as a function of system parameters~\cite{b004, b0010,b0011,b0022,b0023,b0024}. Such stochastic-geometry-based dimensioning frameworks provide tractable alternatives to purely simulation-driven or optimization-heavy planning methods.
\begin{figure}[t]
\centering
\begin{tikzpicture}

\draw[thick] (0,0) circle (2.8);
\node at (0,3.1) {Cell radius $R$};

\draw[dashed] (0,0) circle (1.0);
\draw[dashed] (0,0) circle (1.9);
\node at (1.3,0.3) {\footnotesize Annuli};

\foreach \x/\y in {0.6/0.2, -0.5/0.8, 1.2/-0.9, -1.1/-0.4, 0.2/-1.3} {
  \filldraw (\x,\y) circle (2pt);
}

\node at (0,-3.3) {PPP user locations $\Rightarrow$ spatially varying demand};

\end{tikzpicture}
\caption{Stochastic-geometry-based network dimensioning model. Randomly located users generate spatially dependent subcarrier demand, leading to a random aggregate load whose tail probability characterizes overload events.}
\label{fig:dimensioning}
\end{figure}

\subsection{Multi-class Spatial Model via Poisson Point Processes}
Consider a single downlink cell modeled as a disk $B(0,R)\subset \mathbb{R}^2$. Users are partitioned into $I$ classes corresponding to different IoT services, traffic profiles, or 5G numerologies~\cite{b0013,b0014,b0018}. Class-$i$ users form an independent Poisson point process (PPP) $\Phi_i$ on $B(0,R)$ with intensity $\lambda_i$, a modeling assumption widely adopted in stochastic wireless network analysis~\cite{b0012,b0016,b0017}. By the superposition theorem for PPPs, the union $\Phi = \cup_i \Phi_i$ is also a PPP with total intensity $\lambda=\sum_i \lambda_i$~\cite{b0017}.

\subsection{Subcarrier Demand as a Function of Location and Class}
Let class $i$ be assigned subcarrier bandwidth $W_i$ and require downlink throughput $C_i$, as motivated by 5G New Radio numerology and IoT service differentiation~\cite{b0013,b0014,b0018}. For a user at location $x$, the received SNR under a standard large-scale fading model is
\begin{equation}
\mathrm{SNR}(x) = \frac{P_t K \bar{g}}{n \|x\|^\gamma},
\end{equation}
where $P_t$ is the BS transmit power per user, $K$ is a propagation constant, $\bar{g}$ is the mean shadowing gain, $n$ is noise power, and $\gamma$ is the path-loss exponent~\cite{b003,b0011}. 

For clarity, inter-cell interference is neglected (noise-limited approximation), as commonly done in analytical dimensioning studies~\cite{b0010,b0011,b0022}, although interference-aware extensions exist in the literature~\cite{b005,b0015}.

The number of subcarriers demanded by a class-$i$ user at $x$ is then determined by Shannon capacity:
\begin{equation}
N_i(x)=
\begin{cases}
\left\lceil\frac{C_i}{W_i \log_2\!\left(1+\mathrm{SNR}(x)\right)}\right\rceil, & \mathrm{SNR}(x)\ge \beta_{\min},\\
0, & \text{otherwise,}
\end{cases}
\label{eq:Ni_def}
\end{equation}
where $\beta_{\min}$ is a minimum SNR threshold ensuring coverage and reliable decoding~\cite{b0011,b0022}. The maximum per-user demand for class $i$ is
\begin{equation}
N_{i,\max}=\left\lceil\frac{C_i}{W_i \log_2(1+\beta_{\min})}\right\rceil,
\qquad
N_{\max}=\max_i N_{i,\max}.
\end{equation}

\subsection{Overload Event and Loss Probability}
Let $N_{\mathrm{av}}$ denote the number of available subcarriers at the BS. The total demanded subcarriers in the cell is the PPP functional
\begin{equation}
S \triangleq \sum_{x\in \Phi} N(x)
= \sum_{i} \sum_{x\in \Phi_i} N_i(x).
\end{equation}
A \emph{loss} or \emph{overload} event occurs when demand exceeds supply:
\begin{equation}
P_{\mathrm{loss}} = \mathbb{P}\left(S \ge N_{\mathrm{av}}\right).
\label{eq:Ploss_def}
\end{equation}
Exact evaluation of \eqref{eq:Ploss_def} is possible via combinatorial or convolution-based approaches~\cite{b0010,b0023}, but it becomes increasingly complex for multi-class systems and realistic rate mappings, motivating the use of approximations and bounds~\cite{b0011,b0024}.

\subsection{Concentration-Inequality Upper Bound for Dimensioning}
Using concentration inequalities for functionals of PPPs, explicit upper bounds on the tail probability of $S$ can be derived~\cite{b0010,b0011,b0017}. Let
\begin{equation}
m \triangleq \mathbb{E}[S], \qquad v \triangleq \mathrm{Var}(S),
\end{equation}
which follow directly from Campbell's theorem~\cite{b0012,b0017}. If the add-one cost is bounded as $|D_x S|\le N_{\max}$, then for any $\alpha>1$,
\begin{equation}
\mathbb{P}\left(S \ge \alpha m\right)
\le
\exp\!\left(
-\frac{v}{N_{\max}^2}\,
g\!\left(\frac{(\alpha-1)m N_{\max}}{v}\right)
\right),
\label{eq:ppp_conc_bound}
\end{equation}
where $g(u)=(1+u)\ln(1+u)-u$~\cite{b17}. This bound has been successfully applied to LTE, WiMAX, and OFDMA-based IoT systems for conservative yet tractable dimensioning~\cite{b0010,b0011,b0021}.

\subsection{Closed-Form Computation of $m$ and $v$ via Radial Partitioning}
Because $N_i(x)$ depends only on $\|x\|$ under isotropic path-loss and mean shadowing, it is piecewise constant over annular regions~\cite{b10,b11}. Defining radii $\{R_{i,j}\}$ such that $N_i(x)=j$ for $\|x\|\in [R_{i,j-1},R_{i,j})$, with $R_{i,0}=0$,
\begin{equation}
R_{i,j}=
\left(
\frac{P_t K \bar{g}}{n\left(2^{C_i/(jW_i)}-1\right)}
\right)^{1/\gamma},
\end{equation}
the mean and variance follow in closed form as
\begin{align}
m
&=
\pi \sum_i \lambda_i \sum_{j=1}^{N_{i,\max}}
j\left( (R_{i,j}^2\wedge R^2) - (R_{i,j-1}^2\wedge R^2)\right),\\
v
&=
\pi \sum_i \lambda_i \sum_{j=1}^{N_{i,\max}}
j^2\left( (R_{i,j}^2\wedge R^2) - (R_{i,j-1}^2\wedge R^2)\right),
\end{align}
as previously derived for OFDMA dimensioning problems~\cite{b004}.

\subsection{Empirical Tightness and Design Insights}
Prior numerical and simulation studies indicate that concentration-based bounds are particularly accurate at moderate-to-high user intensities, while being conservative at very low loads~\cite{b0010,b0011,b0023}. The effects of user intensity, class composition, and path-loss exponent on bound tightness have also been investigated in LTE, WiMAX, and IoT-oriented OFDMA systems~\cite{b0021,b0022,b0024}.

Numerical evaluation indicates that the concentration upper bound is \emph{more accurate} (closer to the exact loss probability) at \emph{moderate-to-high user intensities}, while it can be conservative at very low intensities~\cite{b004}. Sensitivity studies further suggest:
\begin{itemize}
\item Increasing overall intensity $\lambda$ generally tightens the bound relative to the exact loss probability.
\item Varying $\alpha$ over a practical range (e.g., $[1.2,2]$) can provide robust dimensioning knobs.
\item Under low user intensity, changes in path-loss exponent $\gamma$ have limited influence on bound tightness, whereas at higher intensity the bound tends to improve with larger $\gamma$ (stronger distance attenuation).
\item For fixed total intensity, redistributing load among classes (changing a single $\lambda_i$ while holding $\sum_i \lambda_i$ fixed) has limited impact on the bound quality in typical regimes.
\end{itemize}
\subsection{Takeaway}
Overall, stochastic-geometry-based dimensioning provides a scalable and analytically grounded alternative to optimization-only or simulation-heavy planning approaches~\cite{b001,b005,b008,b009,b0025}. By translating spatial randomness into resource-demand randomness and controlling overload probability via concentration bounds, the framework enables practical subcarrier dimensioning in multi-class IoT-OFDMA systems consistent with 5G standards~\cite{b004,b0013,b0014,b0018}.

\section{Cell-Free Massive MIMO and Fronthaul-Aware User Management}
\label{sec:cellfree}

Cell-free massive MIMO is a leading 5G/5G-Advanced distributed-RAN paradigm in which a large number of geographically distributed access points (APs) coherently serve a smaller number of user equipments (UEs) through a central processing unit (CPU)~\cite{001b0, 001b1}. While the ``fully cell-free'' idealization assumes that each AP serves all users~\cite{001b1}, practical deployments face computational limits, signaling overhead, and (critically) fronthaul capacity constraints~\cite{001b0,001b10,001b11,001b14}. These constraints motivate \emph{user-centric} (UC) or \emph{user management} designs, where each AP serves only a subset of users, typically selected using large-scale fading statistics~\cite{001b0,001b5,001b6,001b15}.

\begin{figure}[t]
\centering
\begin{tikzpicture}[
    font=\small,
    ap/.style={rectangle, draw, rounded corners, minimum width=0.9cm, minimum height=0.5cm, align=center},
    ue/.style={circle, draw, minimum size=0.55cm, align=center},
    cpu/.style={rectangle, draw, thick, rounded corners, minimum width=1.8cm, minimum height=0.7cm, align=center},
    fr/.style={-, thin},
    serv/.style={-, thick},
    title/.style={font=\bfseries},
    note/.style={font=\scriptsize, align=center}
]

\def\xL{-2.2}
\def\xR{ 2.2}
\def\yCPU{2.15}
\def\yAP{1.15}
\def\yUE{0.0}
\def\dx{1.05}

\node[note] at (0,3.5) {Same physical deployment; different serving sparsity};

\node[title] at (\xL,3.0) {Fully Cell-Free};
\node[cpu] (cpuA) at (\xL,\yCPU) {CPU};

\node[ap] (a1) at (\xL-\dx,\yAP) {AP1};
\node[ap] (a2) at (\xL,\yAP)     {AP2};
\node[ap] (a3) at (\xL+\dx,\yAP) {AP3};

\node[ue] (u1) at (\xL-\dx,\yUE) {U1};
\node[ue] (u2) at (\xL,\yUE)     {U2};
\node[ue] (u3) at (\xL+\dx,\yUE) {U3};

\draw[fr] (a1) -- (cpuA);
\draw[fr] (a2) -- (cpuA);
\draw[fr] (a3) -- (cpuA);

\foreach \apn in {a1,a2,a3}{
  \foreach \uen in {u1,u2,u3}{
    \draw[serv] (\apn) -- (\uen);
  }
}

\node[note] at (\xL,-0.9) {Dense serving graph\\($x_{mk}=1$)};

\node[title] at (\xR,3.0) {User-Centric (UC)};
\node[cpu] (cpuB) at (\xR,\yCPU) {CPU};

\node[ap] (b1) at (\xR-\dx,\yAP) {AP1};
\node[ap] (b2) at (\xR,\yAP)     {AP2};
\node[ap] (b3) at (\xR+\dx,\yAP) {AP3};

\node[ue] (v1) at (\xR-\dx,\yUE) {U1};
\node[ue] (v2) at (\xR,\yUE)     {U2};
\node[ue] (v3) at (\xR+\dx,\yUE) {U3};

\draw[fr] (b1) -- (cpuB) node[midway, left]  {\scriptsize $C_1$};
\draw[fr] (b2) -- (cpuB) node[midway, right] {\scriptsize $C_2$};
\draw[fr] (b3) -- (cpuB) node[midway, right] {\scriptsize $C_3$};

\draw[serv] (b1) -- (v1);
\draw[serv] (b1) -- (v2);
\draw[serv] (b2) -- (v2);
\draw[serv] (b3) -- (v2);
\draw[serv] (b3) -- (v3);

\node[note] at (\xR,-0.9) {Sparse serving graph\\($\sum_k x_{mk}=N_{\mathrm{CUE}}$)};

\end{tikzpicture}
\caption{Fully cell-free and user-centric (UC) architectures under the same physical deployment. UC can be interpreted as a sparsified serving graph of the fully cell-free model, reducing signaling and fronthaul load under per-AP capacity constraints $C_m$.}
\label{fig:cellfree_uc}
\end{figure}
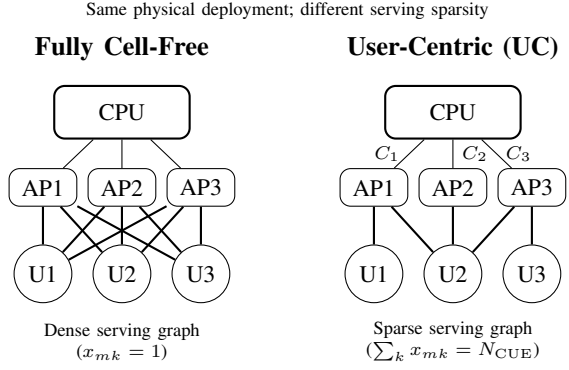

\subsection{System Model and User Management Constraint}
Consider an uplink cell-free network with $M$ single-antenna APs and $K$ single-antenna users. The AP--CPU link for AP $m$ has limited fronthaul capacity $C_m$ (bits/s/Hz)~\cite{001b10,001b11,001b14}. Channels follow
\begin{equation}
g_{mk}=\sqrt{\beta_{mk}}h_{mk}, \qquad h_{mk}\sim\mathcal{CN}(0,1),
\end{equation}
where $\beta_{mk}$ models large-scale fading.

User management is commonly represented using binary association variables $x_{mk}\in\{0,1\}$ indicating whether AP $m$ serves user $k$~\cite{001b5,001b15}. A typical UC-style constraint is
\begin{equation}
\sum_{k=1}^{K}x_{mk}=\text{NCUE}, \qquad \sum_{m=1}^{M}x_{mk}\ge 1,
\label{eq:uc_constraints}
\end{equation}
where NCUE is the target number of served users per AP and the second constraint prevents unserved users~\cite{001b15}.

\paragraph{Heuristic selection with ``no-unserved-user'' safeguard.}
A practical, low-complexity approach is to (i) guarantee each user is connected to at least one AP (e.g., by assigning each user to a nearby AP that has available user slots), and then (ii) fill remaining AP slots with users having the largest $\beta_{mk}$ (large-scale channel gains)~\cite{001b15}. This design retains the core benefits of UC operation, reduced overhead and focused fronthaul usage, while preventing coverage holes that can appear in naive ``strongest-users-only'' rules~\cite{001b15}.

\subsection{Fronthaul-Limited Processing: CFE, ECF, and EMCF}
Limited fronthaul creates a fundamental design choice: whether channel estimation is done at the CPU or at APs, and at what stage compression/quantization is applied~\cite{001b10,001b11,001b14}. Three representative strategies are widely studied:

\paragraph{CFE: Compress--Forward--Estimate.}
APs quantize received pilots and data signals and forward them to the CPU; channel estimation and combining are performed centrally~\cite{001b14}. Since quantization is applied before estimation, pilot quantization noise can degrade estimation quality at the CPU~\cite{001b14}.

\paragraph{ECF: Estimate--Compress--Forward.}
APs estimate channels locally from pilots, quantize the channel estimates (and/or user data streams), and forward only UC-relevant information according to $\{x_{mk}\}$~\cite{001b14}. This enables selective fronthaul usage, often improving performance under tight fronthaul constraints~\cite{001b10,001b11,001b14}.

\paragraph{EMCF: Estimate--Multiply--Compress--Forward.}
APs locally estimate channels, apply a local linear operation (e.g., multiply received data by conjugated channel estimates to form per-user sufficient statistics), and then quantize and forward only the UC-selected components~\cite{001b14}. EMCF can significantly reduce fronthaul load by avoiding raw-signal forwarding~\cite{001b14,001b13}.

From a survey perspective, these strategies share a common structure: the CPU forms per-user soft estimates by combining UC-selected AP contributions (e.g., MRC-style combining for CFE/ECF, or MMSE combining on compressed sufficient statistics for EMCF)~\cite{001b10,001b12,001b14}. In all cases, limited fronthaul manifests as quantization distortion terms that depend on how $C_m$ is allocated between pilot-side and data-side transmissions~\cite{001b10,001b11,001b14}.

\subsection{Impact of User Management Under Limited Fronthaul}
User management changes the fronthaul--performance trade-off because it reshapes how the per-AP fronthaul budget is shared across users~\cite{001b0,001b10,001b11,001b14}:
\begin{itemize}
\item \textbf{ECF/EMCF benefit from limiting NCUE.} When NCUE is small, fewer user streams (or sufficient statistics) are forwarded per AP, increasing the effective number of fronthaul bits per forwarded stream. This reduces quantization noise and can improve both sum-SE and per-user SE~\cite{001b0,001b10,001b11,001b14}.
\item \textbf{CFE is less helped by limiting NCUE.} In CFE, the AP forwards quantized received signals (and pilots), which may still consume fronthaul regardless of which users are later combined at the CPU. As a result, limiting NCUE can reduce macro-diversity gains without a commensurate reduction in fronthaul load, yielding smaller or even negative net gains~\cite{001b0,001b14}.
\item \textbf{Coverage-aware association matters most at low NCUE.} When each AP serves very few users, naive strongest-user selection can leave some users unserved; enforcing the constraint in \eqref{eq:uc_constraints} improves worst-user performance and stabilizes fairness metrics~\cite{001b0,001b15,001b7}.
\end{itemize}

\subsection{Practical Design Guidance}
A practical design pattern emerging from fronthaul-limited UC cell-free studies is~\cite{001b0,001b10,001b11,001b14,001b15}:
\begin{enumerate}
\item Use a coverage-aware heuristic to ensure $\sum_m x_{mk}\ge 1$ for all $k$~\cite{001b0}.
\item Tune NCUE to balance macro-diversity (higher NCUE) versus quantization quality and fronthaul sharing (lower NCUE)~\cite{001b0,001b5,001b15}.
\item Prefer ECF/EMCF-style splits when fronthaul is a dominant bottleneck; use CFE primarily when fronthaul is sufficiently provisioned or when centralized estimation is mandated~\cite{001b0, 001b14,001b10,001b11}.
\end{enumerate}

\subsection{Summary Table}
Table~\ref{tab:user_mgmt_fh} summarizes how user management interacts with fronthaul-limited processing.

\begin{table}[t]
\centering
\caption{Interaction between user management (NCUE) and fronthaul-limited strategies.}
\label{tab:user_mgmt_fh}
\begin{tabular}{p{1.6cm}p{2.2cm}p{2.8cm}}
\toprule
\textbf{Strategy} & \textbf{What is forwarded} & \textbf{Effect of smaller NCUE (typical)} \\
\midrule
CFE & Quantized pilots + raw data signals & Limited benefit; may lose macro-diversity with little fronthaul relief \\
ECF & Quantized channel estimates (and/or selected data) & Often improves (less fronthaul sharing $\Rightarrow$ less quantization noise) \\
EMCF & Quantized per-user sufficient statistics & Often improves; strong fronthaul savings with UC selection \\
\bottomrule
\end{tabular}
\end{table}

\section{Power Allocation and Resource Optimization in Cell-Free 5G Networks}
\label{sec:power}

Power control is a central mechanism in 5G and beyond-5G distributed architectures since it directly mediates the throughput-fairness trade-off under strong inter-user interference, heterogeneous large-scale fading, and practical transport constraints~\cite{b0000,b0001,b0002}. In cell-free and user-centric (UC) massive MIMO, the power allocation problem is further coupled with fronthaul limitations, because the effective noise seen at the central processing unit (CPU) includes both thermal noise and fronthaul-induced quantization distortion~\cite{b00010,b00011,001b14}.

\subsection{Uplink UC Cell-Free Model Under Limited Fronthaul}
Consider an uplink UC cell-free massive MIMO system with $M$ single-antenna access points (APs) serving $K$ single-antenna users, following the UC paradigm introduced in~\cite{b0004,b0005,b0007}. Each AP is connected to the CPU through a fronthaul link with capacity $C_m$ (bits/s/Hz), which limits the amount of pilot and data information that can be conveyed~\cite{b00010,b00012}. Denote the channel between AP $m$ and user $k$ by
\begin{equation}
g_{mk} = \sqrt{\beta_{mk}}h_{mk},
\end{equation}
where $\beta_{mk}$ captures large-scale fading and $h_{mk}\sim \mathcal{CN}(0,1)$ is small-scale fading. 

A UC association is represented by binary variables $x_{mk}\in\{0,1\}$ indicating whether AP $m$ serves user $k$, as commonly adopted in UC cell-free architectures~\cite{b0006,b0007,001b0}. A typical UC constraint enforces that each AP serves a fixed number of users (NCUE), while ensuring every user is served by at least one AP:
\begin{equation}
\sum_{k=1}^{K} x_{mk}=\text{NCUE}, \quad \sum_{m=1}^{M}x_{mk}\ge 1.
\end{equation}
Let $\eta_k\in[0,1]$ denote the uplink power-control coefficient for user $k$, and $\rho_u$ be the nominal uplink SNR scaling, as in~\cite{001b0,b0007,b0000}.

\subsection{Impact of Fronthaul Signaling on Power Allocation}

The fronthaul signaling strategies introduced in the user association discussion
(Compress--Forward--Estimate (CFE) and Estimate--Compress--Forward (ECF))
have a direct and nontrivial impact on power allocation design in cell-free and
user-centric massive MIMO systems.

Under the CFE strategy, power allocation must account for quantization noise
introduced by forwarding raw pilot and data signals to the CPU, as well as
potential fronthaul consumption from users that are not ultimately served.
This coupling often leads to conservative power control policies in fronthaul-
limited regimes~\cite{b0000,b00010,b00011}.

In contrast, the ECF strategy enables more selective power allocation by allowing
APs to forward only UC-relevant channel estimates and data streams.
As a result, transmit power can be more efficiently adapted to both channel
conditions and fronthaul constraints, at the cost of quantization distortion
in locally estimated channels~\cite{b00012,001b14,b0000}.

Overall, the choice of fronthaul signaling strategy fundamentally shapes the
structure and feasibility of power allocation algorithms, highlighting the
strong coupling between physical-layer power control and fronthaul-aware
network design.

From a survey viewpoint, both strategies lead to an effective uplink SINR of the generic form reported in~\cite{b0000}:
\begin{figure*}[t]
\begin{equation}
\mathrm{SINR}_k =
\frac{\rho_u\,\eta_k\!\left(\sum_{m=1}^{M} x_{mk}\,\omega_{mk}\right)^2}
{\rho_u \sum_{k'=1}^{K}\eta_{k'}\!\left(\sum_{m=1}^{M} x_{mk}\,\omega_{mk}\,\beta_{mk'}\right)
+\sum_{m=1}^{M} x_{mk}\,(N+Q_{d,m})\,\omega_{mk}}
\label{eq:generic_sinr}
\end{equation}
\end{figure*}

where $\omega_{mk}$ captures the effective channel-estimation quality term (e.g., $\gamma_{mk}$ or its quantized variant), $N$ is thermal noise power, and $Q_{d,m}$ denotes data-quantization distortion determined by the fronthaul split and allocation of $C_m$ between pilot and data forwarding~\cite{b0000,b00010,001b14}. The main conceptual difference is that in CFE, the CPU-side estimation quality is impacted by pilot quantization, whereas in ECF, the estimation is local but forwarded estimates may be quantized~\cite{b0000,b00012,001b14}.

\subsection{Power Allocation Objectives in UC Cell-Free Networks}
Two canonical power-control objectives frequently used in the 5G and cell-free literature are (i) sum spectral efficiency (sum-SE) maximization and (ii) max--min fairness (minimum-SE maximization)~\cite{b0000,b0006,b0007,b00010}. Under the SINR model in \eqref{eq:generic_sinr}, these can be written as:

\subsubsection{Sum-SE Maximization}
\begin{equation}
\max_{\{\eta_k\}} \quad \sum_{k=1}^{K}\log_2\!\left(1+\mathrm{SINR}_k\right)
\quad \text{s.t.}\quad 0\le \eta_k \le 1.
\label{eq:sumse}
\end{equation}
This objective prioritizes throughput but may disadvantage cell-edge users unless additional fairness mechanisms are included~\cite{b0006,b0007}.

\subsubsection{Minimum-SE (Max--Min) Maximization}
\begin{equation}
\max_{\{\eta_k\}} \quad \min_{k}\;\log_2\!\left(1+\mathrm{SINR}_k\right)
\quad \text{s.t.}\quad 0\le \eta_k \le 1.
\label{eq:maxmin}
\end{equation}
This objective enforces user fairness and is often used as a benchmark for ``uniform service'' in cell-free deployments~\cite{b0000,b00010,b00011,001b14}.

\subsection{Representative Solution Approaches and Practical Couplings}
In practice, power control is coupled with (i) UC association $\{x_{mk}\}$ and (ii) fronthaul partitioning between pilot and data transport (e.g., $C_{p,m}$ and $C_{d,m}$ with $C_{p,m}+C_{d,m}=C_m$)~\cite{b00010,001b14,001b0}. Joint optimization over all variables is typically intractable at scale; hence, a common design pattern is:
\begin{enumerate}
\item Fix UC association using a heuristic based on large-scale fading (serve strongest users per AP while ensuring coverage)~\cite{b0005,b0007,001b0}.
\item Allocate fronthaul budget between pilot and data via low-dimensional search or rules-of-thumb~\cite{001b14,b0000}.
\item Solve power control \eqref{eq:sumse} or \eqref{eq:maxmin} using convex approximations, geometric programming, or linear-programming reformulations under suitable SINR transformations~\cite{b0006,b00010,b00011,b0000}.
\end{enumerate}

\subsection{Design Insights}
A consistent qualitative insight reported across fronthaul-limited UC cell-free studies~\cite{b0000,b00010,b00012,001b14,001b0} is that the effectiveness of power control depends strongly on the fronthaul split:
\begin{itemize}
\item In ECF-type splits, selective forwarding and quantization-aware modeling can make power control more impactful because the CPU receives more ``relevant'' information per fronthaul bit~\cite{b0000,b00012,001b14,001b0}.
\item In CFE-type splits, forwarding the full quantized received signals can make performance more fronthaul-distortion-limited; thus, power control may yield smaller gains compared to ECF in certain regimes~\cite{b0000,b00010,b00011,001b14}.
\end{itemize}

\subsection{Summary Table}
Table~\ref{tab:cfe_ecf_power} summarizes the role of fronthaul in the two strategies and its interaction with power control.

\begin{table}[t]
\centering
\caption{Fronthaul-aware uplink processing and power control in UC cell-free massive MIMO.}
\label{tab:cfe_ecf_power}
\begin{tabular}{p{2.0cm}p{2.2cm}p{2.5cm}}
\toprule
\textbf{Strategy} & \textbf{Where CSI is estimated} & \textbf{Power control sensitivity (typical)} \\
\midrule
CFE & CPU (from quantized pilots) & Often limited by fronthaul distortion; gains may be moderate \\
ECF & AP (local estimation, then quantize/forward) & Often higher sensitivity; gains can be stronger with selective forwarding \\
\bottomrule
\end{tabular}
\end{table}

\section{Open Challenges and Future Research Directions}
\label{sec:future}

Despite significant progress in 5G technologies, several fundamental challenges remain unresolved, particularly when spectrum coexistence, distributed architectures, and resource constraints are considered jointly.
This section highlights open research directions that are critical for the evolution toward 5G-Advanced and future wireless systems~\cite{dahlman2022_5gadv,3gpp_38874}.

\subsection{Joint Optimization Problems}

A major open challenge in 5G systems is the \emph{joint optimization} of power allocation, fronthaul capacity, and user association.
Most existing works decompose these variables and optimize them sequentially or heuristically due to the inherent non-convexity and large-scale nature of the problem~\cite{future_joint1,future_joint2,buzzi2022_cfvision,ngo2023_cfnext}.
However, such decoupled designs may lead to suboptimal performance, especially in cell-free and user-centric massive MIMO architectures where user association directly affects fronthaul load, interference structure, and achievable spectral efficiency.

Joint optimization is further complicated by heterogeneous fronthaul technologies (fiber, microwave, wireless), time-varying channel conditions, and mixed service requirements across eMBB, mMTC, and URLLC.
Developing scalable optimization frameworks that can jointly adapt association, power, and fronthaul resource allocation, while maintaining computational tractability, remains an open research problem.

\subsection{Learning-Based Approaches}

The increasing complexity of 5G networks has motivated the use of machine learning (ML) and data-driven optimization techniques.
Deep learning and reinforcement learning have been proposed for power control, spectrum access, and user association, particularly in scenarios where accurate system models are unavailable or too complex to exploit analytically~\cite{future_ml1,future_ml2,zhang2022_ainative,you2023_6gvision}.

Nevertheless, several challenges limit the practical adoption of learning-based methods.
These include generalization across deployment scenarios, robustness to distribution shifts, interpretability of learned policies, and the need for large training datasets.
In addition, the integration of domain knowledge such as stochastic geometry, physical-layer models, and fronthaul constraints into learning architectures remains an active area of research.
Hybrid approaches that combine model-based optimization with learning-driven adaptation appear especially promising.

\subsection{Mobility and Dynamic Traffic}

Most analytical models for coexistence, dimensioning, and cell-free massive MIMO assume static user locations and stationary traffic patterns.
In practice, user mobility, bursty traffic, and rapidly changing service demands introduce additional layers of complexity~\cite{future_mob1,xiao2021_ntn}.
Mobility impacts channel coherence, association stability, and fronthaul signaling overhead, while dynamic traffic patterns challenge static dimensioning assumptions.

Accurately modeling mobility-aware interference, dynamic association switching, and time-varying resource demand remains an open challenge.
Future research must incorporate spatio-temporal user dynamics into stochastic-geometry-based planning and distributed optimization frameworks to better reflect real-world deployments.

\subsection{Toward 5G-Advanced and 6G}

The evolution toward 5G-Advanced and 6G is expected to further intensify the challenges discussed in this survey.
Emerging directions include ultra-dense cell-free architectures, integrated sensing and communication, reconfigurable intelligent surfaces, sub-THz spectrum usage, and AI-native network control~\cite{future_6g1,future_6g2,liu2022_isac,bjornson2024_6gcf}.

In this context, spectrum coexistence will extend beyond traditional communication systems to include sensing, satellite, and non-terrestrial networks.
Likewise, fronthaul constraints will evolve into broader \emph{compute, communication, and control} bottlenecks.
Developing unified theoretical frameworks that can jointly address coexistence, distributed architectures, and intelligent optimization across these emerging dimensions represents a key research opportunity for the next generation of wireless systems.

\section{Conclusion}
\label{sec:conclusion}

This survey reviewed key technologies and challenges in 5G wireless networks, with emphasis on coexistence, network planning, cell-free architectures, and resource optimization.
The presented discussion highlights the importance of integrated design approaches for future wireless systems.

\bibliographystyle{IEEEtran}
\bibliography{references}

\end{document}